\documentclass{article}

\usepackage[numbers]{natbib}         % bibliography style
\usepackage[colorlinks]{hyperref}    % better urls in bibliography
\usepackage[english]{babel} %%% 'french', 'german', 'spanish', 'danish', etc.
\usepackage{amssymb}
\usepackage{amsmath}
\usepackage{txfonts}
\usepackage{mathdots}
\usepackage[classicReIm]{kpfonts}
\usepackage[dvips]{graphicx} %%% use 'pdftex' instead of 'dvips' for PDF output
\usepackage[a4paper, portrait, margin=1in]{geometry}
\usepackage{tabularx}
\begin{document}

%\selectlanguage{english} %%% remove comment delimiter ('%') and select language if required

\textbf{\Large Deep Learning in Medical Image Registration: A Review}

\textbf{ }

Yabo Fu${}^{1}$, Yang Lei${}^{1}$, Tonghe Wang${}^{1,2}$, Walter J. Curran${}^{1,2}$, Tian Liu${}^{1,2}$ and Xiaofeng Yang${}^{1,2}$*

${}^{1}$Department of Radiation Oncology, Emory University, Atlanta, GA 

${}^{2}$Winship Cancer Institute, Emory University, Atlanta, GA

\noindent 
\bigbreak
\bigbreak
\bigbreak

\textbf{Corresponding author: }

Xiaofeng Yang, PhD

Department of Radiation Oncology

Emory University School of Medicine

1365 Clifton Road NE

Atlanta, GA 30322

E-mail: xiaofeng.yang@emory.edu

\bigbreak
\bigbreak
\bigbreak
\bigbreak
\bigbreak
\bigbreak

\textbf{Abstract}

This paper presents a review of deep learning (DL) based medical image registration methods. We summarized the latest developments and applications of DL-based registration methods in the medical field. These methods were classified into seven categories according to their methods, functions and popularity. A detailed review of each category was presented, highlighting important contributions and identifying specific challenges. A short assessment was presented following the detailed review of each category to summarize its achievements and future potentials. We provided a comprehensive comparison among DL-based methods for lung and brain registration using benchmark datasets. Lastly, we analyzed the statistics of all the cited works from various aspects, revealing the popularity and future trend of DL-based medical image registration.

\noindent \eject 

\noindent 
\section{ Introduction}

Image registration, also known as image fusion or image matching, is the process of aligning two or more images based on image appearances. Medical image registration seeks to find an optimal spatial transformation that best aligns the underlying anatomical structures. Medical image registration is used in many clinical applications such as image guidance \cite{RN124, RN133, RN125, RN209}, motion tracking \cite{RN128, RN126, RN210}, segmentation \cite{RN129, RN130, RN214, RN213, RN211, RN212}, dose accumulation \cite{RN132, RN131}, image reconstruction \cite{RN134} and so on. Medical image registration is a broad topic which can be grouped from various perspectives. From input image point of view, registration methods can be divided into unimodal, multimodal, interpatient, intra-patient (\textit{e.g.} same- or different-day) registration. From deformation model point of view, registration methods can be divided in to rigid, affine and deformable methods. From region of interest (ROI) perspective, registration methods can be grouped according to anatomical sites such as brain, lung registration and so on. From image pair dimension perspective, registration methods can be divided into 3D to 3D, 3D to 2D and 2D to 2D/3D. 

Different applications and registration methods face different challenges. For multi-modal image registration, it is difficult to design an accurate image similarity measures due to the inherent appearance differences between different imaging modalities. Inter-patient registration can be tricky since the underlying anatomical structures are different across patients. Different-day intra-patient registration is challenging due to image appearance changes caused by metabolic processes, bowel movement, patient gaining/losing weight and so on. It is crucial for the registration to be computationally efficient in order to provide real-time image guidance. Examples of such application include 3D-MR to 2D/3D-US prostate registration to guide brachytherapy catheter placement and 3D-CT to 2D X-ray registration in intraoperative surgeries. For segmentation and dose accumulation, it is important to ensure the registration has high spatial accuracy. Motion tracking can be used for motion management in radiotherapy such as patient-setup and treatment planning. Motion tracking could also be used to assess respiratory function through 4D-CT lung registration and to access cardiac function through myocardial tissue tracking. In addition, motion tracking could be used to compensate for irregular motion in image reconstruction. In terms of deformation model, rigid transformation is often too simple to represent the actual tissue deformation while free-form transformation is ill-conditioned and hard to regularize. One limitation of 2D-2D registration is it ignores the out-of-plane deformation. Nevertheless, 3D-3D registration is usually computationally demanding, resulting in slow registration.

Many methods have been proposed to deal with the above-mentioned challenges. Popular registration methods include optical flow \cite{RN140, RN135}, demons \cite{RN143}, ANTs \cite{RN144}, HAMMER \cite{RN146}, ELASTIX \cite{RN147} and so on. Scale invariant feature transform (SIFT) and mutual information (MI) have been proposed for multi-modal image similarity calculation \cite{RN148}. For 3D image registration, GPU has been adopted to accelerate the computational speed \cite{RN149}. Multiple transformation regularization methods including spatial smoothing \cite{RN137}, diffeomorphic \cite{RN143}, spline-based \cite{RN150}, FE-based \cite{RN151} and other deformable models have been proposed. Though medical image registration has been extensively studied, it remains a hot research topic. The field of medical image registration has been evolving rapidly with hundreds of papers published each year. Recently, DL-based methods have changed the landscape of medical image processing research and achieved the-state-of-art performances in many applications\cite{RN158, RN165, RN153, RN159, RN156, RN155, RN163, RN164, RN160, RN166, RN162, RN215, RN167, RN161, RN154, RN152}. However, deep learning in medical image registration has not been extensively studied until the past three to four years. Though several review papers on deep learning in medical image analysis have been published \cite{RN61, RN92, RN58, RN14, RN55, RN10, RN84, RN4}, there are very few review papers that are specific to deep learning in medical image registration \cite{RN26}. The goal of this paper is to summarize the latest developments, challenges and trends in DL-based medical image registration methods. With this survey, we aim to

\begin{enumerate}
\item  Summarize the latest developments in DL-based medical image registration.

\item  Highlight contributions, identify challenges and outline future trends.

\item  Provide detailed statistics on recent publications from different perspectives.
\end{enumerate}

\noindent 

\noindent 
\section{Deep Learning}

\noindent 
\subsection{Convolutional Neural Network}

Convolutional neural network (CNN) is a class of deep neural networks with regularized multilayer perceptron. CNN uses convolution operation in place of general matrix multiplication in simple neural networks. The convolutional filters and operations in CNN make it suitable for visual imagery signal processing. Because of its excellent feature extraction ability, CNN is one of the most successful models for image analysis. Since the breakthrough of AlexNet \cite{RN123}, many variants of CNN have been proposed and have achieved the-state-of-art performances in various image processing tasks. A typical CNN usually consists of multiple convolutional layers, max pooling layers, batch normalization layers, dropout layers, a sigmoid or softmax layer. In each convolutional layer, multiple channels of feature maps were extracted by sliding trainable convolutional kernels across the input feature maps. Hierarchical features with high-level abstraction are extracted using multiple convolutional layers. These feature maps usually go through multiple fully connected layer before reaching the final decision layer. Max pooling layers are often used to reduce the image sizes and to promote spatial invariance of the network. Batch normalization is used to reduce internal covariate shift among the training samples. Weight regularization and dropout layers are used to alleviate data overfitting. The loss function is defined as the difference between the predicted and the target output. CNN is usually trained by minimizing the loss via gradient back propagation using optimization methods. 
Many different types of network architectures have been proposed to improve the performance of CNN \cite{RN92}. U-Net proposed by Ronneberger \textit{et al.} is among one of the most used network architectures \cite{RN168}. U-Net was originally used to perform neuronal structures segmentation. U-Net adopts symmetrical contractive and expansive paths with skip connections between them. U-Net allows effective feature learning from a small number of training datasets. Later, He \textit{et al.} proposed a residual network (ResNet) to ease the difficulty of training deep neural networks \cite{RN169}. The difficulty in training deep networks is caused by gradient degradation and vanishing. They reformulated the layers as learning residual functions instead of directly fitting a desired underlying mapping. Inspired by residual network, Huang \textit{et al.} later proposed a densely connected convolutional network (DenseNet) by connecting each layer to every other layer \cite{RN170}. Inception module was first used in GoogLeNet to alleviate the problem of gradient vanishing and allow for more efficient computation of deeper networks \cite{RN171}. Instead of performing convolution using a kernel with fixed size, an inception module uses multiple kernels of different sizes. The resulting feature maps were concatenated and processed by the next layer. Recently, attention gate was used in CNN to improve performance in image classification and segmentation \cite{RN172}. Attention gate could learn to suppress irrelevant features and highlight salient features useful for a specific task.  

\noindent 
\subsection{Autoencoder}

An autoencoder (AE) is a type of neural network that learns to copy its input to its output without supervision \cite{RN173}. Autoencoder usually consists of an encoder which encodes the input into a low-dimensional latent state space and a decoder which restore the original input from the low-dimensional latent space. To prevent an autoencoder from learning an identity function, regularized autoencoders were invented. Examples of regularized autoencoders include sparse autoencoder, denoising autoencoder and contractive autoencoder \cite{RN174}. Recently, convolutional autoencoder (CAE) was proposed to combine CNN with traditional autoencoders \cite{RN175}. CAE replaces the fully connected layer in traditional AE with convolutional layers and transpose-convolutional layers. CAE has been used in multiple medical image processing tasks such as lesion detection, segmentation, image restoration \cite{RN92}. Different from above-mentioned AEs, variational AE (VAE) is generative model that learns latent representation using a variational approach \cite{RN176}. VAE has been used for anomaly detection \cite{RN177} and image generation \cite{RN178}.  

\noindent 
\subsection{Recurrent Neural Network}

A recurrent neural network (RNN) is a type of neural network that was used to model dynamic temporal behavior \cite{RN179}. RNN is widely used for natural language processing \cite{RN180}. Unlike feedforward networks such as CNN, RNN is suitable for processing temporal signal. The internal state of RNN was used to model and ‘memorize’ previously processed information. Therefore, the output of RNN was dependent on not only its immediate input but also its input history. Long short-term memory (LSTM) is one type of RNN which has been used in image processing tasks \cite{RN181}. Recently, Cho \textit{et al.} proposed a simplified version of LSTM, called gated recurrent unit \cite{RN182}. 

\noindent 
\subsection{Reinforcement Learning}

Reinforcement learning (RL) is a type of machine learning that focused on predicting the best actions to take given its current state in an environment \cite{RN183}. RL is usually modeled as a Markov decision process using a set of environment states and actions. An artificial agent is trained to maximize its cumulative expected rewards. The training process often involves an exploration-exploitation tradeoff. Exploration means to explore the whole space to gather more information while exploitation means to explore the promising areas given current information. Q-learning is a model-free RL algorithm, which aims to learn a Q function that models the action-reward relationship. Bellman equation is often used in Q-learning for reward calculation. The Bellman equation calculates the maximum future reward as the immediate reward the agent gets for entering the current state plus a weighted maximum future reward for the next state. For image processing, the Q function is often modeled as CNN, which could encode input images as states and learn the Q function via supervised training \cite{RN110, RN95, RN93, RN89}. 

\noindent 
\subsection{Generative Adversarial Network}

A typical generative adversarial network (GAN) consists of two competing networks, a generator and a discriminator \cite{RN184}. The generator is trained to generate artificial data that approximate a target data distribution from a low-dimensional latent space. The discriminator is trained to distinguish the artificial data from actual data. The discriminator encourages the generator to predict realistic data by penalizing unrealistic predictions via learning. Therefore, the discriminative loss could be considered as a dynamic network-based loss term. The generator and discriminator both are getting better during training to reach Nash equilibrium. Multiple variants of GAN include conditional GAN (cGan) \cite{RN185}, InfoGan \cite{RN186}, , CycleGAN \cite{RN188}, StarGan \cite{RN187} and so on. In medical imaging, GAN has been used to perform image synthesis for inter- or intra-modality, such as MR to synthetic CT \cite{RN156, RN160}, CT to synthetic MR \cite{RN165, RN216}, CBCT to synthetic CT \cite{RN159}, non-attenuation correction (non-AC) PET to CT \cite{RN217}, low-dose PET to synthetic full-dose PET \cite{RN164}, non-AC PET to AC PET \cite{RN218}, low-dose CT to full-dose CT \cite{RN219} and so on. In medical image registration, GAN is usually used to either provide additional regularization or translate multi-modal registration to unimodal registration. Out of medical imaging, GAN has been widely used in many other fields including science, art, games and so on. 

\noindent 

\noindent 
\section{Deep learning in medical image registration}

DL-based registration methods can be classified according to deep learning properties, such as network architectures (CNN, RL, GAN etc.), training process (supervised, unsupervised etc.), inference types (iterative, one-shot prediction), input image sizes (patch-based, whole image-based), output types (dense transformation, sparse transformation on control points, parametric regression of transformation model etc.) and so on. In this paper, we classified DL-based medical image registration methods according to its methods, functions and popularity in to seven categories, including 1) RL-based methods, 2) Deep similarity-based methods, 3) Supervised transformation predication, 4) Unsupervised transformation prediction, 5) GAN in medical image registration, 6) Registration validation using deep learning, and 7) Other learning-based methods. In each category, we provided a comprehensive table, listing all the surveyed works belonging to this category and summarizing their important features. 

Before we delve into the details of each category, we provided a detailed overview of DL-based medical image registration methods with their corresponding components and features in Fig. 1. The purpose of Fig. 1 is to give the readers an overall understanding of each category by putting its important features side by side with each other. CNN was initially designed to process highly structured datasets such as images, which are usually expressed by regular grid-sampling data points. Therefore, almost all cited methods have utilized convolutional kernels in their deep learning design. This explains why the CNN module is in the middle of Fig. 1.

\begin{figure}
\centering
\noindent \includegraphics*[width=6.50in, height=4.20in, keepaspectratio=true]{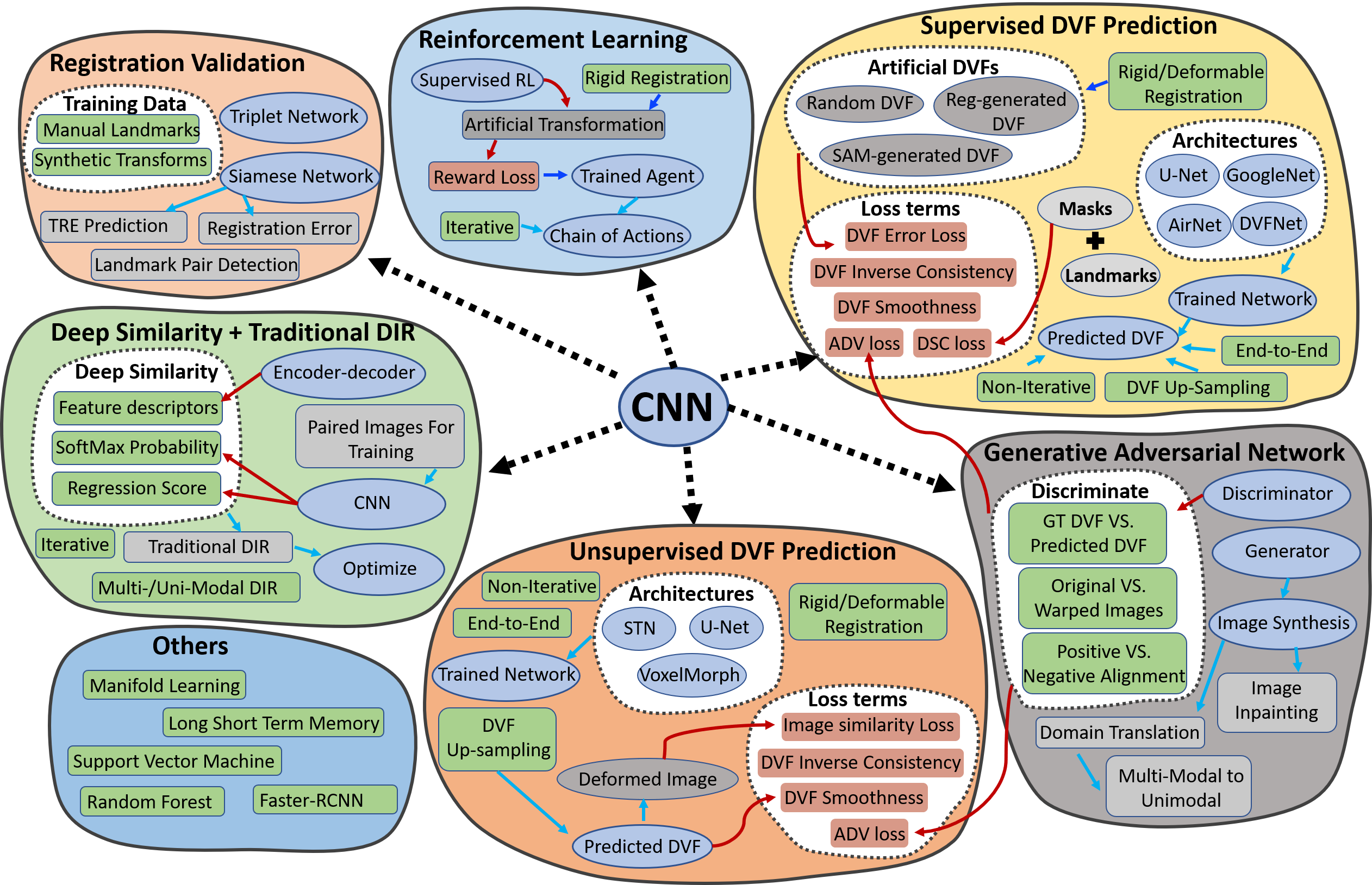}

\noindent Fig. 1. Overview of seven categories of DL-based methods in medical image registration
\end{figure}

\noindent 

Works cited in this review were collected from various databases, including Google Scholar, PubMed, Web of Science, Semantic Scholar and so on. To collect as many works as possible, we used a variety of keywords including but not limited to machine learning, deep learning, learning-based, convolutional neural network, image registration, image fusion, image alignment, registration validation, registration error prediction, motion tracking, motion management and so on. We totally collected over 150 papers that are closely related to deep learning-based medical image registration. Most of these works were published between the year of 2016 and 2019. The number of publications is plotted against year by stacked bar charts in Fig. 2. Number of papers were counted by categories. The total number of publications has grown dramatically over the last few years. Fig. 2 shows a clear trend of increasing interest in supervised transformation prediction (SupCNN) and unsupervised transform prediction (UnsupCNN). Meanwhile, GAN are gradually gaining popularity. On the other hand, the number of papers of RL-based medical image registration has decreased in 2019, which may indicate decreasing interest in RL for medical image registration. The `DeepSimilarity' in Fig. 2 represents the category of deep similarity-based registration methods. The number of papers in this category has also increased, however, only slightly as compared to `SupCNN' and `UnsupCNN' categories. In addition, more and more studies were published on using DL for medical image registration validations.

\begin{figure}
\centering
\noindent \includegraphics*[width=4.77in, height=3.00in, keepaspectratio=true]{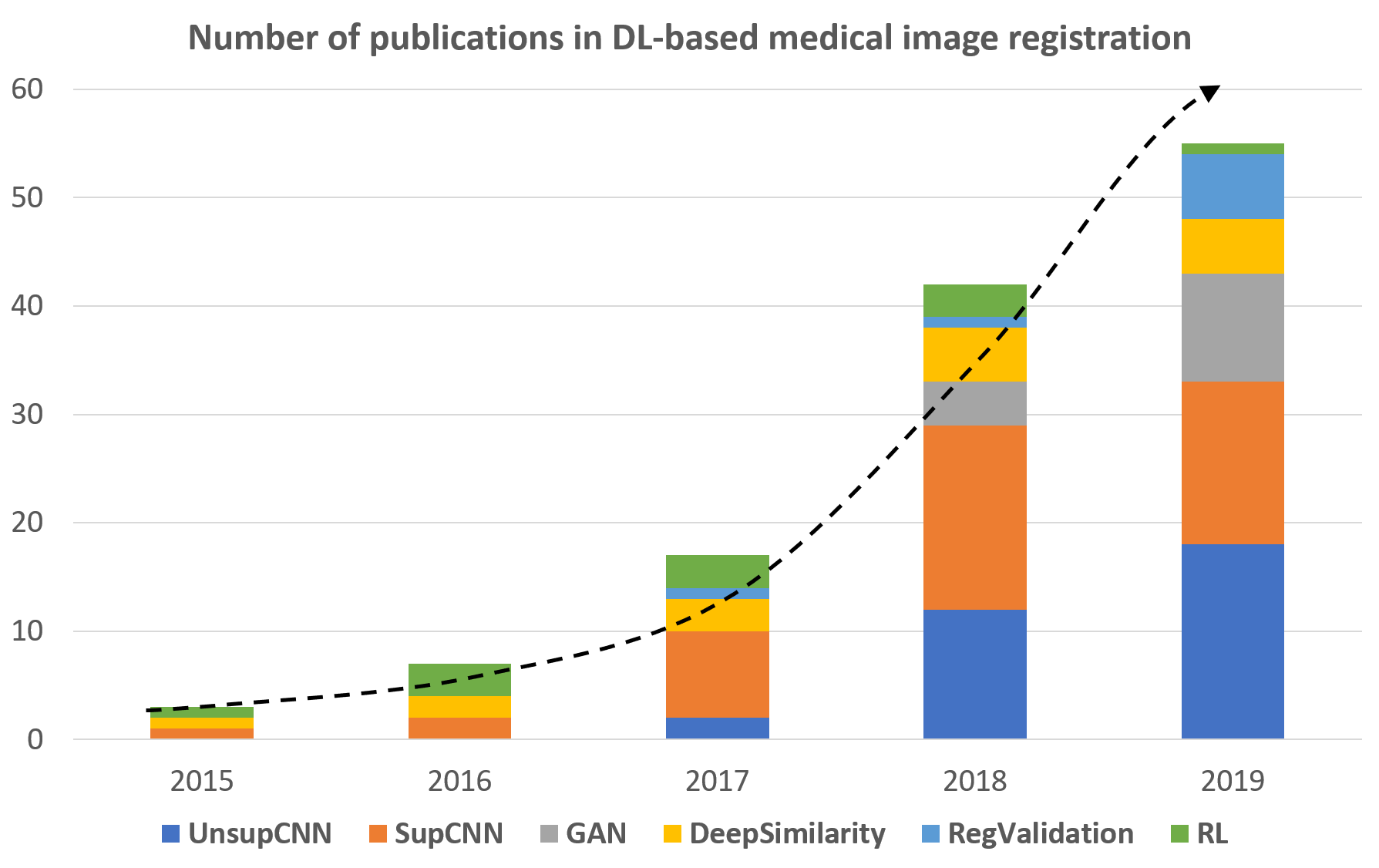}

\noindent Fig. 2. Overview of number of publications in DL-based medical image registration. The dotted line indicates increase interest in DL-based registration methods over the years. `DeepSimilarity' is the category of using DL-based similarity measures in traditional registration frameworks. `RegValidation' represents the category of using DL for registration validation.
\end{figure}

\noindent 
\subsection{Deep similarity-based methods}
Conventional intensity-based similarity metrics include sum-of-square distance (SSD), mean square distance (MSD), (normalized) cross correlation (CC), and (normalized) mutual information (MI). Generally, conventional similarity measures work quite well for unimodal image registration where the image pair shares the same intensity distribution such as CT-CT, MR-MR image registration. However, noise and artifacts in images such as US and CBCT often cause conventional similarity measures to perform poorly even in unimodal image registration. Metrics such as SSD and MSD does not work for multi-modal image registration. To develop a similarity measure for multi-modality image registration, handcrafted descriptors such as MI were proposed. To improve its performance, a variety of MI variants such as correlation ratio-based MI \cite{RN15}, contextual conditioned MI \cite{RN3000} and modality independent neighborhood descriptor (MIND) \cite{RN114} have been proposed. Recently, CNN has achieved huge success in tasks such as image classification and segmentation problems. However, CNN has not been widely used in image registration tasks until the last three to four years. To take the advantage of CNN, several groups tried to replace the traditional image similarity measures such as SSD, MAE and MI with DL-based similarity measures, achieving promising registration results. In the following section, we described several important works that attempted to use DL-based similarity measures in medical image registration. 

\bigbreak

\noindent 
{\bf 3.1.1 Overview of works}

Cheng \textit{et al.} proposed a deep similarity learning network to train a binary classifier \cite{RN118}. The network was trained to learn the correspondence of two image patches from CT-MR image pair. The continuous probabilistic value was used as the similarity score. Similarly, Simonovsky \textit{et al.} proposed a 3D similarity network using a few aligned image pairs \cite{RN106}. The network was trained to classify whether an image pair is aligned or not. They observed that hinge loss performed better than cross entropy. The learnt deep similarity metric was then used to replace MI in traditional deformable image registration (DIR) for brain T1-T2 registration. It is important to ensure the smoothness of first order derivative in order to fit the deep similarity metrics into traditional DIR frameworks. The gradient of the deep similarity metric with respect to transformation was calculated using chain rule. They found out that high overlap of neighboring patches led to smoother and more stable derivatives. They have trained the network using IXI brain datasets and tested it using a completely independent datasets called ALBERTs in order to show the good generality of the learnt metric. They showed that the learnt deep similarity metric outperformed MI by a significant margin.

\begin{table}
\small
\centering
\textbf{Table 1} Overview of deep similarity-based methods

\begin{tabular*}{\textwidth}{c @{\extracolsep{\fill}} ccccc} \\ \hline 
\textbf{References} & \textbf{ROI} & \textbf{Dimension} & \textbf{Modality} & \textbf{Transformation} & \textbf{Supervision} \\ \hline 
\cite{RN106} & Brain & 3D-3D & T1-T2 & Deformable & Supervised \\ \hline 
\cite{RN103} & Brain & 3D-3D & MR & Deformable & Unsupervised \\ \hline 
\cite{RN12} & Brain & 3D-3D & MR & Rigid, Deformable & Supervised \\ \hline 
\cite{RN118} & Brain & 2D-2D & MR-CT & Rigid & Supervised \\ \hline 
\cite{RN13} & Prostate & 3D-3D & MR-US & Rigid & Supervised \\ \hline 
\cite{RN50} & Brain & 2D-2D & MR & Rigid & Weakly Supervised \\ \hline 
\cite{RN11} & Brain, HN, Abdomen & 3D-3D & MR, CT & Deformable & Weakly Supervised \\ \hline 
\end{tabular*}
\end{table}
\bigbreak

Compared to CT-MR and T1-T2 image registration, MR-US image registration is more challenging due to the fundamental imaging acquisition differences between MR and US. A deep learning-based similarity measure is desired for MR-US image registration. Haskins \textit{et al.} proposed to use CNN to predict the target registration error (TRE) between 3D MR and transrectal US (TRUS) images \cite{RN13}. The predicted TRE was used as image similarity metric for MR-US rigid registration. TREs obtained from expert-aligned images were used as ground truth. The CNN was trained to regress to the TRE as similarity prediction. The learnt metric was non-smooth and non-convex, which hinders gradient-based optimization. To address this issue, they performed multiple TRE predictions throughout the optimization. The average TRE estimate was used as the similarity metric to mitigate the non-convex problem and to expand the capture range. They claimed that the learnt similarity metric outperformed MI and its variant MIND \cite{RN114}.

In previous works, accurate image alignment is needed for deep similarity metrics learning. However, it is very difficult to obtain well aligned multi-modal image pairs for network training. The quality of image alignment could affect the accuracy of the learnt deep similarity metrics. To mitigate this problem, Sedghi \textit{et al.} used special data augmentation techniques called dithering and symmetrizing to discharge the need for well-aligned images for deep metric learning \cite{RN50}. The learnt deep metric outperformed MI on 2D brain image registration. Though they managed to relax the absolute accuracy of image alignment in network training, roughly-aligned image pairs were still necessary. To eliminate the need for aligned image pairs, Wu \textit{et al.} proposed to use stacked autoencoders (SAE) to learn intrinsic feature representations by unsupervised learning \cite{RN103}. The convolutional SAE could encode an image to obtain low-dimensional feature representations for image similarity calculation. The learnt feature representations were used in Demons and HAMMER to perform brain image DIR. They showed that the image registration performance has improved consistently using the learnt feature representations in terms of dice similarity coefficient (DSC). To test the generality of the learnt feature representation, they reused network trained using LONI dataset on ADNI datasets. The results were comparable to the case of learning feature representation from the same datasets. 

It was shown that combining multi-metric measures could produce more robust registration results compared to using the metrics individually. Ferrante \textit{et al.} used support vector machine (SVM) to learn weights of an aggregation of similarity measures including anatomical segmentation maps and displacement vector labels \cite{RN11}. They have showed that the multi-metric outperformed conventional single-metric approaches. To deal with the non-convex of the aggregated similarity metric, they optimized a regularized upper bound of the loss using CCCP algorithm \cite{RN198}. One limitation of this method was that segmentation masks of the source images were needed at testing stage.

\bigbreak

\noindent
{\bf 3.1.2 Assessments}

Deep similarity metric has shown its potential to outperform traditional similarity metrics in medical image registration. However, it is difficult to ensure that its derivative is smooth for optimization. The above-mentioned measures of using a large overlap \cite{RN106} or performing multiple TRE predictions \cite{RN13} are computationally demanding and only mitigate the problem of non-convex derivatives. Well-aligned image pairs are difficult to obtain for deep similarity network training. Though  Wu \textit{et al.} \cite{RN103} has demonstrated that deep similarity network could be trained in an unsupervised manner, they only tested on unimodal image registration. Extra experiments on multi-modal images need to be performed to show its effectiveness. The biggest limitation of this category maybe that the registration process still inherits the iterative nature of traditional DIR frameworks, which slows the registration process. As more and more papers on direct transformation prediction emerge, it is expected that this category will be less attractive in the future. 

\noindent 
\subsection{Reinforcement learning in medical image registration}

One disadvantage of the previous category is that the registration process is iterative and time-consuming. It is desired to develop a method to predict transformation in one shot. However, one shot transformation prediction is very difficult due to the high dimensionality of the output parameter space. RL has recently gained a lot of attention since the publications from Mnih \textit{et al.} \cite{RN199} and Silver \textit{et al.} \cite{RN200}. They combined RL with DNN to achieve human-level performances on Atari and Go. Inspired by the success of RL, and to circumvent the challenge of high dimensionality in one shot transformation prediction, several groups proposed to combine CNN with RL to decompose the registration task into a sequence of classification problems. The strategy is to find a series of actions, such as rotation and translation along certain axis by a certain value, to iteratively improve image alignment. 

\noindent

\begin{table}
\small
\centering
\textbf{Table 2} Overview of RL in medical image registration

\begin{tabular*}{\textwidth}{c @{\extracolsep{\fill}} cccc} \\ \hline 
\textbf{References} & \textbf{ROI} & \textbf{Dimension} & \textbf{Modality} & \textbf{Transformation} \\ \hline 
\cite{RN189, RN110} & Cardiac, HN & 2D, 3D & MR, CT, US & NA \\ \hline 
\cite{RN95} & Prostate & 3D-3D & MR & Deformable \\ \hline 
\cite{RN93} & Spine, Cardiac & 3D-3D & CT-CBCT & Rigid \\ \hline 
\cite{RN190} & Chest, Abdomen & 2D-2D & CT-Depth Image & Rigid \\ \hline 
\cite{RN89} & Spine & 3D-2D & 3DCT-Xray & Rigid \\ \hline 
\cite{RN37} & Spine & 3D-2D & 3DCT-Xray & Rigid \\ \hline 
\cite{RN121} & Nasopharyngeal & 2D-2D & MR-CT & Rigid with scaling \\ \hline 
\end{tabular*}
\end{table}
\bigbreak

\bigbreak

\noindent
{\bf 3.2.1 Overview of works}

Table 2 shows a list of selected references that used RL in medical image registration. Liao \textit{et al.} was one of the first to explore RL in medical image registration \cite{RN93}. The task was to perform 3D-3D, rigid, cone beam CT (CBCT)-CT image registration. Specific challenges of the registration include large differences in field of views (FOVs) between the CT and CBCT in spine registration and the severe streaking artifacts in CBCT. An artificial agent was trained using a greedy supervised approach to perform rigid image registration. The artificial agent was modelled using CNN, which took raw images as input and output the next optimal action. The action space consists of 12 candidate transformations, which are ±1mm of translation and ±1 degree of rotation along the x, y, and z axis, respectively. Ground truth alignment were obtained using iterative closest point registration of expert-defined spine landmarks and epicardium segmentation, followed by visual inspection and manual editing. Data augmentation was used to artificially de-align the image pair with known transformations. Different from Mnih \textit{et al.} who trained their network with repeated trial and error, Liao \textit{et al.} trained the network with greedy supervision, where the reward can be calculated explicitly via a recursive function. They showed that the network training process with supervision was a magnitude more efficient than the training process of Mnih \textit{et al.}’s network. They also claimed their network could reliably overcome local maxima, which was challenging for generic optimization algorithms when the underlying problem was non-convex.
Motivated by \cite{RN93}, Miao \textit{et al.} proposed a multi-agent system with an auto attention mechanism to rigidly register 3D-CT with 2D X-ray spine image \cite{RN89}. Reliable 2D-3D image registration could map the pre-operative 3D data to real-time 2D X-ray images by image fusion. To deal with various image artifacts, they proposed to use an auto-attention mechanism to detect regions with reliable visual cues to drive the registration. In addition, they used a dilated FCN-based training mechanism to reduce the degree of freedom of training data to improve the training efficiency. They have outperformed single agent-based and optimization-based methods in terms of TRE.
Sun \textit{et al.} proposed to use an asynchronous RL algorithm with customized reward function for 2D MR-CT image registration \cite{RN121}. They used datasets from 99 patients diagnosed as nasopharyngeal carcinoma. Ground truth image alignments were obtained using toolbox Elastix \cite{RN147}. Different from previous works, Sun \textit{et al.} incorporated scaling factor into the action space. The action space consists of 8 candidate transformations including ±1 pixel for translation, ±1 degree for rotation and ±0.05 for scaling. CNN was used to encode image states and LSTM was used to encode hidden states between neighboring frames. Their method was better than Elastix in terms of TRE when the initial image alignment was poor. The use of actor-critic scheme \cite{RN201} allowed the agent to explore transformation parameter spaces freely and avoided local minima when the initial alignment was poor. On the contrary, when the initial image alignment was good, Elastix was slightly better than their method. In the inference phase, a Monte Carlo rollout strategy was proposed to terminate the searching path to reach a better action. 
All of the above-mentioned methods focused on rigid registration since rigid transformation could be represented by a low-dimensional parametric space, such as rotation, translation and scaling. However, non-rigid, free-form transformation model has high dimensionality and non-linearity which would result in a huge action space. To deal with this problem, Krebs \textit{et al.} proposed to build a statistical deformation model (SDM) with a low-dimensional parametric space \cite{RN95}. Principal component analysis (PCA) was used to construct SDM on B-spline deformation vector field (DVF). Modes of the PCA of the displacement were used as the unknow vectors for the agents to optimize. They evaluated the method on inter-subject MR prostate image registration in both 2D and 3D. The method achieved DSC scores of 0.87 and 0.80 for 2D and 3D, respectively.
Ghesu \textit{et al.} proposed to use RL to detect 3D-landmarks in medical images \cite{RN110}. This method was mentioned since it belongs to the category of RL and the detected landmarks could be used for landmark-based image registration. They reformulated the landmark detection task as a behavioral problem for the network to learn. To deal with local minima problem, a multi-scale approach was used. Experiments on 3D-CT scans were conducted to compare with another five methods. The results showed that the detection accuracy was improved by 20-30 percent while being 2-3 orders of magnitude faster. 

\bigbreak

\noindent 
{\bf 3.2.2 Assessment}

The biggest limitation of RL-based image registration is that the transformation model is highly constrained to low-dimensionality. As a result, most of the RL-based registration methods used rigid transformation models. Though Krebs \textit{et al.} has applied RL to non-rigid image registration by predicting a low-dimensional parametric space of statistical deformation model, the accuracy and flexibility of the deformation model is highly constrained and may not be adequate to represent the actual deformation. RL-based image registration methods have shown its usefulness in enhancing the robustness of many algorithms in multi-modal image registration tasks. Despite the usefulness of RL, statistics indicates loss of popularity of this category, evidenced by the decreasing number of papers in 2019. As the techniques advance, more and more direct transformation predication methods are proposed. The accuracy of the direct transformation prediction methods is constantly improving, achieving comparable accuracy to top traditional DIR methods. Therefore, the advantage of casting registration as a sequence of classification problems in RL-based registration methods is gradually vanishing. 

\noindent 

\noindent 
\subsection{Supervised transformation predication}

\begin{table}
\small
\centering
\textbf{Table 3} Overview of supervised transformation prediction methods

\begin{tabular*}{\textwidth}{c @{\extracolsep{\fill}} ccccc} \\ \hline 
\textbf{References} & \textbf{ROI} & \textbf{Dimension} & \textbf{Patch-based} & \textbf{Modality} & \textbf{Transformation} \\ \hline 
\cite{RN108} & Implant, TEE & 2D-3D & Yes & Xray & Rigid \\ \hline 
\cite{RN87} & Cranial & 2D-3D & No & CBCT-Xray & Deformable \\ \hline 
\cite{RN86} & Cardiac & 3D-3D & No & MR & Deformable \\ \hline 
\cite{RN82} & Cardiac, Brain & 2D-2D & No & MR & Deformable \\ \hline 
\cite{RN79} & Brain & 3D-3D & Yes & MR & Deformable \\ \hline 
\cite{RN74} & Brain & 3D-3D & Yes & MR & Deformable \\ \hline 
\cite{RN75} & Pelvic & 3D-3D & Yes & MR-CT & Deformable \\ \hline 
\cite{RN66} & Cardiac & 2D-2D & No & MR & Deformable \\ \hline 
\cite{RN64, RN65} & Prostate & 3D-3D & No & MR-US & Deformable \\ \hline 
\cite{RN57} & Abdomen & 2D-2D & Yes & MR & Deformable \\ \hline 
\cite{RN51} & Brain & 3D-3D/2D & No & MR & Rigid \\ \hline 
\cite{RN49} & Lung & 3D-3D & Yes & CT & Deformable \\ \hline 
\cite{RN46} & Brain & 2D-2D & No & T1-T2 & Rigid \\ \hline 
\cite{RN43} & Liver & 2D-2D & Yes & CT-US & Affine \\ \hline 
\cite{RN39} & Prostate & 3D-3D & No & MR-US & Rigid + Affine \\ \hline 
\cite{RN33, RN69} & Lung & 3D-3D & No & CT & Deformable \\ \hline 
\cite{RN29} & Brain & 3D-3D & Yes & MR & Deformable \\ \hline 
\cite{RN98} & Lung & 3D-2D & No & CT & Deformable \\ \hline 
\cite{RN17} & Brain & 3D-3D & No & T1, T2, Flair & Affine \\ \hline 
\cite{RN16} & Skull, Upper Body & 2D-2D & No & DRR-Xray & Deformable \\ \hline 
\cite{RN53, RN83, RN24} & Lung & 3D-3D & Yes & CT & Deformable \\ \hline 
\end{tabular*}
\end{table}
\bigbreak

Both deep similarity-based and RL-based registration methods are iterative methods in order to avoid the challenges of one-shot transformation prediction. Despite the difficulties, several groups have attempted to train networks to directly infer the final transformation in a single forward prediction. The challenges include 1) high dimensionality of the output parametric space, 2) lack of training datasets with ground truth transformations and 3) regularization of the predicted transformation. Methods including ground truth transformation generation, image re-sampling and transformation regularization methods have been proposed to overcome these challenges. Table 3 shows a list of selected references that used supervised transformation prediction for medical image registration.
\bigbreak

\noindent 
{\bf 3.3.1 Overview of works}

\noindent 
{\bf 3.3.1.1 Ground truth transformation generation}

For supervised transformation prediction, it is important to generate many image pairs with known transformations for network training. Numerous data augmentation techniques were proposed for artificial transformations generation. Generally, these artificial transformation generation methods can be classified into three groups: 1) random transformation, 2) traditional registration-generated transformation and 3) model-based transformations.

\noindent 
{\bf A. Random transformation generation}

Salehi \textit{et al.} aimed to speed up and improve the capture range of 3D-3D and 2D-3D rigid image registration of fetal brain MR scans \cite{RN51}. CNN was used to predict both rotation and translation parameters. The network was trained using datasets generated by randomly rotating and translating the original 3D images. Both MSE and geodesic distance were used for loss function calculation. Geodesic distance is the distance between two points on a unit sphere. They have showed significant improvement after combining the geodesic distance loss with the MSE loss. Sun \textit{et al.} used expert aligned CT-US image pairs as ground truth \cite{RN43}. Known artificial affine transformations were used to synthesize training datasets. The network was trained to predict the affine parameters. They have trained network which worked for simulated CT-US registration. However, it does not work on real CT-US pairs due to the vast appearance differences between the simulated and the real US. They have tried multiple methods to counter-act overfitting, such as deleting dropout layers, less complex network, parameter regularization and weight decay. Unfortunately, none of them worked.
Eppenhof \textit{et al.} proposed to train a CNN using synthetic random transformations to perform 3D-CT lung DIR \cite{RN69}. The output of the network was DVF on a thin plate spline transform grid. MSE between the predicted DVF and the ground truth DVF was used as loss function. They achieved 4.02±3.08 mm TRE on DIRLAB \cite{RN202}, which was much worse than 1.36+1.01 mm of traditional DIR method. They later improved their method to use a U-Net architecture \cite{RN33}. The network was trained on whole image. Images were down-sampled to fit into GPU memory. Again, synthetic random transformation was used to train the network. Affine pre-registration was required prior to CNN transformation prediction. They managed to reduce the TRE from 4.02±3.08 mm to 2.17±1.89 mm on DIRLAB datasets. Despite the slightly worse TRE than traditional DIR methods, they have demonstrated the possibility of direct transformation prediction using CNN.

Eppenhof \textit{et al.} proposed to train a CNN using synthetic random transformations to perform 3D-CT lung DIR [117]. The output of the network was DVF on a thin plate spline transform grid. MSE between the predicted DVF and the ground truth DVF was used as loss function. They achieved 4.02$\mathrm{\pm}$3.08 mm TRE on DIRLAB [126], which was much worse than 1.36+1.01 mm of traditional DIR method. They later improved their method to use a U-Net architecture [118]. The network was trained on whole image. Images were down-sampled to fit into GPU memory. Again, synthetic random transformation was used to train the network. Affine pre-registration was required prior to CNN transformation prediction. They managed to reduce the TRE from 4.02$\mathrm{\pm}$3.08 mm to 2.17$\mathrm{\pm}$1.89 mm on DIRLAB datasets. Despite the slightly worse TRE than traditional DIR methods, they have demonstrated the possibility of direct transformation prediction using CNN. 

\noindent 
{\bf B. Traditional registration-generated transformations}

Later, several groups tried to use traditional registration methods to register an image pair to generate ‘ground truth’ transformations for the network to learn. The rationale is that random transformation generation might be too different from the true transformation, which might deteriorate the performance of network. 
Sentker \textit{et al.} used DVF generated from traditional DIRs including PlastiMatch \cite{RN205}, NiftyReg \cite{RN206} and VarReg \cite{RN207} as ground truth \cite{RN49}. MSE between the predicted and the ground truth DVF was used as loss function to train a network for 3D-CT lung registration. On DIRLAB \cite{RN202} datasets, they achieved better TRE using DVFs generated by VarReg as compared to PlastiMatch and NiftyReg. Results showed that their CNN-based registration method was comparable to the original traditional DIR in terms of TRE. The best TRE values they have achieved on DIRLAB is 2.50±1.16 mm. Fan \textit{et al.} proposed a BIRNet to perform brain image registration using dual supervision \cite{RN29}. Ground truth transformations were obtained using existing registration methods. MSE between the ground truth and the predicted transformations were used as loss function. They used not only the original image but also its difference and gradient images as input to the network.

\noindent 
{\bf C. Model-based transformation generation}

Uzunova \textit{et al.} aimed to generate a large and diverse set of training image pairs with known transformations from a few sample images \cite{RN82}. They proposed to learn highly expressive statistical appearance models (SAM) from a few training samples. Assuming Gaussian distribution for the appearance parameters, they synthesized huge amounts of realistic ground truth training datasets. FlowNet \cite{RN203} architecture was used to register 2D MR cardiac images. For comparison, they have generated ground truth transformations using three different methods, which are affine registration-generated, randomly-generated and the proposed SAM-generated transformations. They showed that CNN learnt from the SAM-generated transformation outperformed CNN learnt from randomly-generated and affine registration-generated transformation.
Sokooti \textit{et al.} generated artificial DVFs using model-based respiratory motion to simulate ground truth DVF for 3D-CT lung image registration \cite{RN24}. For comparison, random transformations were also generated using single frequency and mixed frequencies. They tested different combinations of various network structures including U-Net whole image, multi-view based and U-Net advanced. The multi-view and U-Net advanced all used patch-based training. TRE and Jacobian determinant were used as evaluation metrics. After comparison, they claimed that the realistic model-based transformation performed better compared to random transformations in terms of TRE. On average, they achieved TRE of 2.32 mm and 1.86 mm for SPREAD and DIRLAB datasets, respectively.

\bigbreak

\noindent 
{\bf 3.3.1.2 Supervision methods}

As neural network develops, many new supervision terms such as `supervised', `unsupervised', `deeply supervised', `weakly supervised', `dual supervised', `self-supervised' have emerged. Generally, neural network learns to perform a certain task by minimizing a predefined loss function via optimization. These terms refer to how the training datasets are prepared and how the networks are trained using the datasets. In the following paragraph, we briefly describe the definition of each supervision strategy in the context of DL-based image registration.

 The learning process of a neural network is supervised if the desired output is already known in the training datasets. Supervised network means the network is trained with the ground truth transformation, which is a dense DVF for free deformation and a parametric vector of 6 for rigid transformation. On the other hand, unsupervised learning has no target output available in the training datasets, which means the desired DVFs or target transformation parameters are absent in the training datasets. Unsupervised network was also referred to self-supervised network since the warped image is generated from one of the input image pair and compared to another input image for supervision. Deep supervision usually means that the differences between outputs from multiple layers and the desired outputs are penalized during training whereas normal supervision only penalizes the difference between the final output and the desired output. In this manner, supervision was extended to deep layers of the network. Weak supervision represents scenario where ground truth other than the exact desired output is available in the training datasets and used to calculate the loss function. For example, a network is called weakly supervised if corresponding anatomical structural masks or landmark pairs, not the desired dense DVF, are used to train the network for direct dense DVF prediction. Dual supervision means that the network is trained using both supervised and unsupervised loss functions.

\noindent 
{\bf A. Weak supervision}

Methods that use ground truth transformation generation were mainly supervised method for direct transformation prediction. Weakly supervised transformation prediction has also been explored. Instead of using artificially-generated transformations, Hu \textit{et al.} proposed to use higher-level correspondence information such as labels of anatomical organs for network training \cite{RN64}. They argued that such anatomical labels were more reliable and practical to obtain. They trained a CNN to perform deformable MR-US prostate image registration. The network was trained using weakly supervised method, meaning that only corresponding anatomical labels, not dense voxel-level spatial correspondence, were used for loss calculation. The anatomical labels were required only in the training stage for loss calculation. Labels were not required in inference stage to facilitate fast registration. Similarly, Hering \textit{et al.} combined the complementary information from segmentation labels and image similarity to train a network \cite{RN66}. They showed significant higher DSC scores than using only image similarity loss or segmentation label loss in 2D MR cardiac DIR.

\noindent 
{\bf B. Dual supervision }

Technically, dual supervision is not strictly defined. It usually means the network was trained using two types of important loss functions. Cao \textit{et al.} used dual supervision which includes a MR-MR loss and a CT-CT loss \cite{RN75}. Prior to network training, they transformed the multi-modality to unimodality registration by using pre-aligned counterpart images, for MR-CT registration. The MR has a pre-aligned CT and CT has a pre-aligned MR. The loss function has a dual similarity loss including MR-MR and CT-CT loss. They showed that the dual-modality similarity performed better than SyN \cite{RN204} and single modality similarity in terms of DSC and average surface distance (ASD) in pelvic image registration. Liu \textit{et al.} used representation learning to learn feature-based descriptors with probability maps of confidence level \cite{RN16}. Then, the learnt descriptor pairs across the image were used to build a geometric constraint using Hough voting or RANSAC. The network was trained using both supervised synthetic transformations and an unsupervised descriptor image similarity loss. Similarly, Fan \textit{et al.} combined both supervised and unsupervised loss terms for dual supervision in MRI brains image registration \cite{RN29}.

\noindent 
\bigbreak
{\bf 3.3.2 Assessment}

In recent two to three years, we have seen a huge interest in supervised CNN direct transformation prediction, evidenced by increasing number of publications. Though direct transformation prediction has yet to outperform the-state-of-art traditional DIR methods, the registration accuracy has improved greatly. Some methods have achieved comparable registration accuracy to the traditional DIR methods. Ground truth transformation generation will continue to play an important role in network training. Limitations of using artificially generated image pair with known ground truth transformations include 1) the generated transformation might not reflect the true physiological motion, 2) the generated transformation might not capture the large range of variations of actual image registration scenarios and 3) the artificially generated image pairs in the training stage are different from the actual image pair in the inference stage. To deal with the first limitations, we can use various transformation generation models. Adequate data augmentation could be performed to mitigate the second limitation. Domain adaption \cite{RN67, RN37} could be used to account for the domain difference between the artificially-generated and the true images. Image registration is an ill-posed problem, the ground truth transformation could help to constrain the final transformation prediction. Combinations of different loss functions and DVF regularization methods have also been examined to improve the accuracy of registration. We expect DL-based registration of this category to keep growing in the future.

\noindent 
\subsection{Unsupervised transformation prediction}

It is desired to develop unsupervised image registration methods to overcome the lack of training datasets with known transformations. However, it is difficult to define proper loss function of the network without ground truth transformations. In 2015, Jaderberg \textit{et al.} proposed a spatial transformer network (STN) which explicitly allows spatial manipulation of data within the network \cite{RN111}. Importantly, the spatial transformer network was a differentiable module that can be inserted in to existing CNN architectures. The publication of STN has inspired many unsupervised image registration methods since STN enables image similarity loss calculation during the training process. A typical unsupervised transformation prediction network for DIR takes an image pair as input and directly output dense DVF, which was used by STN to warp the moving image to generate warped images. The warped images were then compared to fixed images to calculate image similarity loss. DVF smoothness constraint was normally used to regularize the predicted DVF.

\bigbreak

\noindent 
{\bf 3.4.1 Overview of works}

\begin{table}
\small
\centering
\textbf{Table 4} Overview of unsupervised transformation prediction methods

\begin{tabular*}{\textwidth}{c @{\extracolsep{\fill}} ccccc} \\ \hline 
\textbf{References} & \textbf{ROI} & \textbf{Dimension} & \textbf{Patch-based} & \textbf{Modality} & \textbf{Transformation} \\ \hline 
\cite{RN96} & Brain & 3D-3D & No & MR & Deformable \\ \hline 
\cite{RN85} & Brain, Liver & 2D-2D & No & MR, CT & Deformable \\ \hline 
\cite{RN81} & Cardiac  & 2D-2D & No & MR & Deformable \\ \hline 
\cite{RN78} & Neural tissue & 2D-2D & No & EM & Deformable \\ \hline 
\cite{RN73} & Brain & 3D-3D & No & MR & Affine \\ \hline 
\cite{RN122, RN30} & Brain & 3D-3D & Yes & MR & Deformable \\ \hline 
\cite{RN67} & Lung, Cardiac & 2D-2D & No & MR, Xray & Deformable \\ \hline 
\cite{RN62} & HN & 3D-3D & Yes & CT & Deformable \\ \hline 
\cite{RN60} & Cardiac & 3D-3D & No & MR & Deformable \\ \hline 
\cite{RN59} & Brain & 3D-3D & No & MR & Deformable \\ \hline 
\cite{RN120, RN52} & Cardiac & 2D-2D & No & MR & Deformable \\ \hline 
\cite{RN48} & Cardiac & 2D-2D & No & MR & Deformable \\ \hline 
\cite{RN191} & Neuron Tissue & 2D-2D & Yes & EM & Affine \\ \hline 
\cite{RN45} & Lung & 3D-3D & No & MR & Deformable \\ \hline 
\cite{RN44} & Brain & 3D-3D & No & MR-US & Deformable \\ \hline 
\cite{RN41} & Cardiac, Lung & 3D-3D & Yes & MR, CT & Affine and Deformable \\ \hline 
\cite{RN38} & Brain & 3D-3D & No & MR & Deformable \\ \hline 
\cite{RN77, RN36, RN71} & Brain & 3D-3D & No & MR & Deformable \\ \hline 
\cite{RN34, RN115} & Prostate & 3D-3D & Yes & CT & Deformable \\ \hline 
\cite{RN30} & Brain, Pelvic & 3D-3D & Yes & MR, CT & Deformable \\ \hline 
\cite{RN25} & Lung & 2D-2D & Yes & CT & Deformable \\ \hline 
\cite{RN97} & Liver & 3D-3D & No & CT & Deformable \\ \hline 
\cite{RN19, RN18} & Brain & 3D-3D & No & MR & Deformable \\ \hline 
\cite{RN20} & Liver & 3D-3D & No & CT & Deformable \\ \hline 
\cite{RN21} & Abdomen & 3D-3D & Yes & CT & Deformable \\ \hline 
\cite{RN22} & Retina & 2D-2D & No & FA & Deformable \\ \hline 
\cite{RN8} & Neural tissue & 2D-2D & No & EM & Deformable \\ \hline 
\cite{RN5} & Abdominopelvic & 3D-3D & Yes & CT-PET & Deformable \\ \hline 
\cite{RN127} & Lung, Cardiac & 3D-3D & Yes & CT, MR & Deformable \\ \hline 
\end{tabular*}
\end{table}
\bigbreak

Yoo \textit{et al.} proposed to use a convolution autoencoder (CAE) to encode image to a vector to calculate similarity, called feature-based similarity which is different from handcrafted feature similarity such as SIFT \cite{RN78}. They showed this feature-based similarity measure was better than intensity-based similarity measure for DIR. They have combined the deep similarity metrics and STN for unsupervised transformation estimation in 2D electron microscopy (EM) neural tissue image registration. Balakrishnan \textit{et al.} proposed an unsupervised CNN-based DIR method for MR brain atlas-based registration \cite{RN77, RN36}. They used a U-Net like architecture and named it ‘VoxelMorph’. In the training, the network penalized the differences in image appearances with the help of STN. Smoothness constraint was used to penalize local spatial variations in the predicted transformation. They have achieved comparable performance to ANT \cite{RN144} registration method in terms of DSC score of multiple anatomical structures. Later, they extended their method to leverage auxiliary segmentations available in the training data. A DSC loss function was added to the original loss functions in the training stage. Segmentation labels were not required during testing. They investigated unsupervised brain registration, with and without segmentation label DSC loss. Their results showed that the segmentation loss could help yield improved DSC scores. The performance is comparable to ANT and NiftyReg, while being x150 faster than ANTs and x40 faster than NiftyReg.
Like \cite{RN36}, Qin \textit{et al.} also used segmentation as complementary information for cardiac MR image registration \cite{RN52}. They found out that the feature learnt by registration CNN could be used in segmentation as well. The predicted DVF was used to deform the masks of moving image to generate masks of the fixed image. They trained a joint segmentation and registration model for cardiac cine image registration and proved that the joint mode could generate better results than the separate models alone in both segmentation and registration tasks. Similar idea has been explored in \cite{RN56} as well. They claimed registration and segmentation are complementary functions and combining them can improve each other’s performance. 
Later, Zhang \textit{et al.} proposed a network with trans-convolutional layers for end-to-end DVF prediction in MR brain DIR \cite{RN38}. They focused on the diffeomorphic mapping of the transformation. To encourage smoothness and avoid folding of the predicted transformation, they proposed an inverse-consistent regularization term to penalize the difference between two transformations from the respective inverse mappings. The loss function consists of an image similarity loss, a transformation smoothness loss, an inverse consistent loss and an anti-folding loss. Their method has outperformed Demons and Syn, in terms of DSC score, sensitivity, positive predictive value, average surface distance and Hausdorff distance. A similar idea was proposed by Kim \textit{et al.} who used cycle consistent loss to enforce DVF regularization \cite{RN97}. They also used identity loss where the output DVF should be zero if the moving and fixed image are the same image.
For 3D-CT image registration, Lei \textit{et al.} used an unsupervised CNN to perform abdominal image registration \cite{RN21}. They used a dilated inception module to extract multi-scale motion features for robust DVF prediction. Apart from the image similarity loss and DVF smoothness loss, they integrated a discriminator to provide additional adversarial loss for DVF regularization. Vos \textit{et al.} proposed an unsupervised affine and DIR framework by stacking multiple CNN into a larger network \cite{RN41}. The network was tested on cardiac cine MRI and 3D CT lung image registration. They showed their method was comparable to conventional DIR method while being several orders of magnitude faster. Like \cite{RN41}, Lau \textit{et al.} cascaded affine and deformable networks for CT liver DIR \cite{RN20}. Recently, Jiang \textit{et al.} proposed a multi-scale framework with unsupervised CNN for 3D CT lung DIR \cite{RN25}. They cascaded three CNN models with each model focusing on its own scale level. The network was trained using image patches to optimize an image similarity loss and a DVF smoothness loss. They showed that network trained on SPARE datasets could generalize to a different DIRLAB datasets. In addition, the same trained network also performed well on CT-CBCT and CBCT-CBCT registration without retraining or fine-tuning. They achieved an average TRE of 1.66±1.44 mm on DIRLAB datasets. Fu \textit{et al.} proposed an unsupervised method for 3D-CT lung DIR \cite{RN192}. They first performed whole-image registration on down-sampled image using a CoarseNet to warp the moving image globally. Then, image patches of the globally warped moving image were registered to the image patches of the fixed image using a patch-based FineNet. They also incorporated a discriminator to provide adversarial loss by penalizing unrealistic warped images. Vessel enhancement was performed prior to DIR to improve the registration accuracy. They have achieved an average TRE of 1.59±1.58 mm, which outperformed some traditional DIR methods. Interestingly, both Jiang \textit{et al.} and Fu \textit{et al.} have achieved better TRE values using unsupervised methods than the supervised methods in \cite{RN33} and \cite{RN49}.

\bigbreak

\noindent 
{\bf 3.4.2 Assessment}

Compared to supervised transformation prediction, unsupervised methods effectively alleviate the problem of lack of training datasets. Various regularization terms have been proposed to encourage plausible transformation prediction. Several groups have achieved comparable or even better results in terms of TRE on DIRLAB 3D-CT lung DIR. However, most of the methods in this category focused on unimodality registration. There has been a lack of investigation in multi-modality image registration using unsupervised methods. To provide additional supervision, several groups have combined supervised with unsupervised methods for transformation prediction \cite{RN29}. The combination seems beneficial; however, more investigation was needed to justify its effectiveness. Given the promising results of the unsupervised methods, we expect a continuous growth of interest in this category.

\noindent 
\subsection{GAN in medical image registration}

The use of GAN in medical image registration can be generally categorized in two groups: 1) to provide additional regularization of the predicted transformation; 2) to perform cross-domain image mapping.

\bigbreak

\noindent 
{\bf 3.5.1 Overview of works}

\begin{table}
\small
\centering
\textbf{Table 5} Overview of registration methods using GAN

\begin{tabular*}{\textwidth}{c @{\extracolsep{\fill}} ccccc} \\ \hline 
\textbf{References} & \textbf{ROI} & \textbf{Dimension} & \textbf{Patch-based} & \textbf{Modality} & \textbf{Transformation} \\ \hline 
\cite{RN122} & Brain & 3D-3D & Yes & MR & Deformable \\ \hline 
\cite{RN65} & Prostate & 3D-3D & No & MR-US & Deformable \\ \hline 
\cite{RN39} & Prostate & 3D-3D & No & MR-US & Deformable \\ \hline 
\cite{RN51} & Brain & 3D-3D & No & MR & Rigid \\ \hline 
\cite{RN115} & Prostate & 3D-3D & Yes & CT & Deformable \\ \hline 
\cite{RN30} & Brain, Pelvic & 3D-3D & Both & MR, CT & Deformable \\ \hline 
\cite{RN21} & Abdomen & 3D-3D & Yes & CT & Deformable \\ \hline 
\cite{RN192} & Lung & 3D-3D & Yes & CT & Deformable \\ \hline 
\cite{RN120, RN56, RN22} & Retina, Cardiac  & 2D-2D & No & FA, Xray & Deformable \\ \hline 
\cite{RN31} & Lung, Brain & 2D-2D & No & T1-T2, CT-MR & Deformable \\ \hline 
\end{tabular*}
\end{table}

\bigbreak

\noindent 
{\bf 3.5.1.1 GAN-based regularization}

Since image registration is an ill-posed problem, it is crucial to have adequate regularization to encourage plausible transformations and to prevent unrealistic transformations such as tissue folding. Commonly used regularization terms include DVF smoothness constraint, anti-folding constraint and inverse consistency constraint. However, it remains ambiguous whether these constraints are adequate for proper regularization. Recently, GAN-based regularization terms have been introduced to the realm of image registration. The idea is to train an adversarial network to introduce a network-based loss for transformation regularization. In the literature, discriminators were trained to distinguish three types of inputs, including 1) whether a transformation is predicted or ground truth, 2) whether an image is realistic or warped by predicted transformation, 3) whether an image pair alignment is positive or negative.
Yan \textit{et al.} trained an adversarial network to tell whether an image was deformed using ground truth transformation or predicted transformation \cite{RN39}. Randomly generated transformations from manually aligned image pairs were used as ground truth to train a network to perform MR-US prostate image registration. The trained discriminator could provide not only an adversarial loss for regularization but also a discriminator score for alignment evaluation. Fan \textit{et al.} used a discriminator to distinguish whether an image pair were well aligned \cite{RN30}. In unimodal image registration, they have defined a positive image alignment case as weighted linear combination of the fixed and the moving images. In multi-modal image registration case, positive image alignments were pre-defined using paired MR and CT images. They performed on MR brain images for unimodal registration and on pelvic CT-MR for multi-modal registration. They have showed that the performance increased with the adversarial loss. Lei \textit{et al.} used a discriminator to judge whether the warped image is realistic enough to the original images \cite{RN21}. Fu \textit{et al.} used a similar idea and showed that the inclusion of adversarial loss could improve registration accuracy in 3D-CT lung DIR \cite{RN192}.
The above GAN-based methods have tried to introduce regularization from the image or transformation appearance perspective. Differently, Hu \textit{et al.} tried to introduce biomechanical constraints to 3D MR-US prostate image registration by discriminating whether a transformation is predicted or generated by finite element analysis \cite{RN65}. Instead of adding the adversarial loss to existing smoothness loss, they replaced the smoothness loss with the adversarial loss. They showed that their method could predict physically plausible deformation without any other smoothness penalty. 

\bigbreak

\noindent 
{\bf 3.5.1.2 GAN-based cross-domain image mapping}

For multi-modal image registration, progresses have been made by using deep similarity metrics in traditional DIR frameworks. Using iterative methods, several works have outperformed the-state-of-art MI similarity measures. However, in terms of direct transformation prediction, multi-modal image registration has not benefited from DL as much as unimodal image registration has. This is mainly due to the vast appearance differences between different modalities. To overcome this challenge, GAN has been used to translate multi-modal to unimodal image registration by mapping images from one modality to another.
Salehi \textit{et al.} trained a CNN using T2-weighted images to perform fetal brain MR registration. They tested the network on T1-weighted images by first mapping the T1 to T2 image domain using a conditional GAN \cite{RN51}. They showed the trained network generalized well on the synthesized T2 images. Qin \textit{et al.} used an unsupervised image-to-image translation framework to cast multi-modal to unimodal image registration \cite{RN31}. The image to image translation method assumes the images could be decomposed into content code and style code. They have showed comparable results to MIND and Elastix on BraTs datasets in terms of RMSE of DVF error. On COPDGene datasets, they outperformed MIND and Elastix in terms of DICE, mean contour distance (MCD) and Hausdorff distance. Mahapatra \textit{et al.} combined cGan \cite{RN185} and registration network together to directly predict both DVF and warped image \cite{RN22}. They implicitly transformed image in one modality to another modality. They outperformed Elastix on 2D retinal image registration in terms of Hausdorff distance, MAD and MSE. Elmahdy \textit{et al.} claimed that inpainting gas pockets in the rectum could enhance rectum and seminal vesicle registration \cite{RN115}. They used GAN to detect and inpaint rectum gas pocket prior to image registration. 

\bigbreak

\noindent 
{\bf 3.5.2 Assessment}

\noindent GAN has been shown to be promising in medical image registration via either novel adversarial loss or image domain translation. For adversarial losses, GAN could provide learnt network-based regularizations that are complementary to traditional handcrafted regularization terms. For image domain translation, GAN effectively cast the more challenging multi-modal registration to unimodal image registration, which allows many existing unimodal registration algorithms to be applied to multi-modal image registration. However, the absolute intensity mapping accuracy of GAN is yet to be investigated. GAN has also been applied to deep similarity metric learning in registration and alignment validation. As evidenced by the trend in Fig. 2, we expect to see more papers using GAN in image registration tasks in the future. 

\noindent 

\noindent 
\subsection{Registration validation using deep learning}

The performance of image registration could be evaluated using image similarity metrics such as SSD, NCC and MI. However, the image similarity metrics only evaluate the overall alignment on the whole image. To have a deeper insight into local registration accuracy, we usually rely on manual landmark pair selection. Nevertheless, manual landmark pair selection is time-consuming, subjective and error-prone especially when many landmarks were to be selected. Fu \textit{et al.} used a Siamese network for large quantity landmark pair detection on 3D-CT lung images \cite{RN28}. The network was trained using the manual landmark pairs from DIRLAB datasets. They performed experiments comparisons, showing that the network could outperform human in landmark pair detection. Neylon \textit{et al.} proposed to use a deep neural network to predict TRE for given image similarity metrics \cite{RN88}. The network was trained using patient-specific biomechanical models of head-neck anatomy. They demonstrated that the network could rapidly and accurately quantify registration performance.

\bigbreak

\noindent 
{\bf 3.6.1 Overview of works}

\begin{table}
\small
\centering
\textbf{Table 6} Overview of registration validation methods using deep learning

\begin{tabular*}{\textwidth}{c @{\extracolsep{\fill}} cccc} \\ \hline 
\textbf{References} & \textbf{ROI} & \textbf{Dimension} & \textbf{Modality} & \textbf{End point} \\ \hline 
\cite{RN88} & HN & 3D & CT & TRE prediction \\ \hline 
\cite{RN70} & Lung & 3D & CT & Registration error \\ \hline 
\cite{RN35} & Brain & 3D & MRI & DSC score \\ \hline 
\cite{RN28} & Lung & 3D & CT & Landmark Pairs \\ \hline 
\cite{RN27} & Lung & 3D & CT & Registration error \\ \hline 
\cite{RN9} & Lung & 3D & CT & Registration error \\ \hline 
\end{tabular*}
\end{table}
\bigbreak

Eppenhof \textit{et al.} proposed a TRE alternative to assess DIR registration accuracy. They used synthetic transformations as ground truth to avoid the need for manual annotations \cite{RN70}. The ground truth error map was the L2 difference between ground truth transformations and the predicted transformations. They trained a network to robustly estimate registration errors with sub-voxel accuracy. Galib \textit{et al.} predicted an overall registration error index, which is the ratio between good alignment sub-volumes and poor alignment sub-volumes \cite{RN27}. They justified the choice of threshold TRE of 3.5mm as a cutoff value of good and bad alignment. Their network was trained using manually labeled landmarks from DIRLAB. Sokooti \textit{et al.} proposed a random forest regression method for quantitative error prediction of DIR \cite{RN9}. They used both intensity-based features such as MIND and registration-based features such as transformation Jacobian determinant. Dubost \textit{et al.} used ventricle DSC score to evaluate brain registration \cite{RN35}. The ventricle was segmented using deep learning-based method. 

\bigbreak

\noindent 
{\bf 3.6.2 Assessment}

The number of papers using deep learning for registration evaluation has increased significantly in 2019. Most works treated registration error prediction as a supervised regression problem. Network was trained using manually annotated datasets. It is important to make sure the ground truth datasets are of high quality. Most of existing methods focused on lung because benchmark datasets with manual landmark pairs exists for 3D CT lung such as DIRLAB. It would be interesting to see the method be applied on many other treatment sites. Unsupervised registration error prediction is another interesting research topic to eliminate the need for manual annotated datasets. 

\noindent 

\noindent 
\subsection{Other learning-based methods in medical image registration}

Jiang \textit{et al.} proposed to use CNN to lean and infer expressive sparse multi-grid configurations prior to B-spline coefficient optimization \cite{RN63}. Liu \textit{et al.} used a ten-layer FCN for image synthesis without GAN to transform multimodal to unimodal registration among T1-weighted, T2-weighed, and proton density images \cite{RN157}. Then, they used Elastix software with SSD similarity metric for the registration of brain phantom and IXI datasets. They outperformed MI similarity index. Wright \textit{et al.} proposed to use LSTM network to predict a rigid transformation and an isotropic scaling factor for MR-US fetal brain registration \cite{RN40}. Bashiri \textit{et al.} used Laplacian eigenmap as a manifold learning method to implement a multi-modal to unimodal image translation in 2D brain image registration \cite{RN32}.

\begin{table}
\small
\centering
\textbf{Table 7} Overview of other deep learning-based image registration methods

\begin{tabular*}{\textwidth}{c @{\extracolsep{\fill}} ccccc} \\ \hline 
\textbf{References} & \textbf{ROI} & \textbf{Dimension} & \textbf{Modality} & \textbf{Transformation} & \textbf{Methods} \\ \hline 
\cite{RN63} & Lung & 3D-3D & CT & Deformable & Multi-grid Inference \\ \hline 
\cite{RN40} & Brain & 3D-3D & MR-US & Rigid & LSTM \\ \hline 
\cite{RN32} & Brain & 2D-2D & CT, T1, T2, PD & Rigid & Manifold Learning \\ \hline 
\cite{RN7} & Brain, Abdomen & 2D-3D & CT-PET, CT-MRI & Deformable & CAE, DSCNN \\ \hline 
\cite{RN6} & Spine & 3D-2D & 3DCT-Xray & Rigid & FasterRCNN \\ \hline 
\cite{RN15} & Brain & 2D-2D & T1-T2, T1-PD & Deformable & FCN \\ \hline 
\cite{RN37}& Spine & 3D-2D & 3DCT-Xray & Rigid & Domain adaptation \\ \hline 
\end{tabular*}
\end{table}
\bigbreak

Yu \textit{et al.} proposed to use FasterRCNN \cite{RN208} for vertebrae bounding box detection \cite{RN6}. The detected bounding box was then matched to doctor-annotated bounding box on the X-ray image. Zheng \textit{et al.} proposed a domain adaptation module to cope with the domain variance between synthetic data and real data \cite{RN37}. The adaptation module can be trained using a few paired real and synthetic data. The trained module could be plugged into the network to transfer the real features to approach the synthetic features. Since network was trained on synthetic data, the network should perform well on synthetic data. Hence, it is reasonable to transfer the real data features to synthetic features. 

\noindent 
\section{Benchmark}

Benchmarking is important for readers to understand through comparison the advantages and disadvantages of each method. For image registration, both registration accuracy and computational time could be benchmarked. However, researchers have been reporting registration accuracies more than the computational speed. Computational speed is largely dependent on the hardware, which is often different from group to group. According to the statistics of the cited works, the top two ROIs of registration are brain and lung. Therefore, we summarized the registration datasets for brain, registration accuracies for lung.

\noindent 
\subsection{Lung}

DIRLAB is one of the most cited public datasets for 4D-CT chest image registration studies \cite{RN202}. DIRLAB provides 300 manually selected landmark pairs for end-exhalation and end-inhalation phases. This dataset was frequently used for 4D-CT lung registration benchmarking. To provide the readers a better understanding of the latest DL-based registration, we have listed the TREs of three top performing traditional methods and seven DL-based lung registration methods. Table 8 shows that DL-based lung registration methods have yet outperformed the top traditional DIR methods. However, DL-based DIR methods have been making substantial improvement over the years, with Fu \textit{et al.} and Jiang \textit{et al.} almost achieving comparable and slightly better TRE than Delmon \textit{et al.} TREs of traditional DIR on case 8 were consistently better than that of the DL-based DIR. Case 8 is one of the most challenging cases in the DIRLAB datasets with impaired image quality and significant lung motion. This phenomenon suggests that the robustness and competency of DL-based DIR need to be further improved.

\begin{table}
\footnotesize
\centering
\textbf{Table 8} Comparison of Target Registration Error (TRE) values among different methods on DIRLAB datasets, TRE unit: (mm), *: Traditional DIR methods

\begin{tabular}{|>{\centering\arraybackslash}m{0.2in}|>{\centering\arraybackslash}m{0.4in}|>{\centering\arraybackslash}m{0.4in}|>{\centering\arraybackslash}m{0.4in}|>{\centering\arraybackslash}m{0.4in}|>{\centering\arraybackslash}m{0.4in}|>{\centering\arraybackslash}m{0.4in}|>{\centering\arraybackslash}m{0.4in}|>{\centering\arraybackslash}m{0.4in}|>{\centering\arraybackslash}m{0.4in}|>{\centering\arraybackslash}m{0.4in}|>{\centering\arraybackslash}m{0.4in}|} \hline 
\textbf{Set} & \textbf{Initial} & \textbf{Heinrich* \textit{et al.}\newline }\cite{RN194}\textbf{} & \textbf{Delmon* \textit{et al.}\newline }\cite{RN193}\textbf{} & \textbf{Staring* \textit{et al.}\newline }\cite{RN195}\textbf{}\textbf{} & \textbf{Eppenhof \textit{et al. }}\cite{RN33}\textbf{} & \textbf{De Vos \textit{et al.}\newline }\cite{RN41}\textbf{} & \textbf{Sentker \textit{et al.} }\cite{RN49}\textbf{} & \textbf{Fu \textit{et al.}\newline }\cite{RN192}\textbf{} & \textbf{Sokooti \textit{et al.} }\cite{RN24}\textbf{} & \textbf{Jiang \textit{et al.} }\cite{RN25}\textbf{} & \textbf{Fechter \textit{et al.} }\cite{RN127}\textbf{} \\ \hline 
1 & 3.89$\mathrm{\pm}$2.78 & 0.97$\mathrm{\pm}$0.5  & 1.2$\mathrm{\pm}$0.6 & 0.99$\mathrm{\pm}$0.57 & 1.45$\mathrm{\pm}$1.06  & 1.27$\mathrm{\pm}$1.16  & 1.20$\mathrm{\pm}$0.60  & 0.98$\mathrm{\pm}$0.54  & 1.13$\mathrm{\pm}$0.51  & 1.20$\mathrm{\pm}$0.63  & 1.21$\mathrm{\pm}$0.88  \\ \hline 
2 & 4.3$\mathrm{\pm}$3.9 & 0.96$\mathrm{\pm}$0.5 & 1.1$\mathrm{\pm}$0.6 & 0.94$\mathrm{\pm}$0.53 & 1.46$\mathrm{\pm}$0.76 & 1.20$\mathrm{\pm}$1.12 & 1.19$\mathrm{\pm}$0.63 & 0.98$\mathrm{\pm}$0.52 & 1.08$\mathrm{\pm}$0.55 & 1.13$\mathrm{\pm}$0.56 & 1.13$\mathrm{\pm}$0.65 \\ \hline 
3 & 6.94$\mathrm{\pm}$4.05 & 1.21$\mathrm{\pm}$0.7  & 1.6$\mathrm{\pm}$0.9 & 1.13$\mathrm{\pm}$0.64 & 1.57$\mathrm{\pm}$1.10  & 1.48$\mathrm{\pm}$1.26  & 1.67$\mathrm{\pm}$0.90  & 1.14$\mathrm{\pm}$0.64  & 1.33$\mathrm{\pm}$0.73  & 1.30$\mathrm{\pm}$0.70  & 1.32$\mathrm{\pm}$0.82  \\ \hline 
4 & 9.83$\mathrm{\pm}$4.86 & 1.39$\mathrm{\pm}$1.0  & 1.6$\mathrm{\pm}$1.1 & 1.49$\mathrm{\pm}$1.01 & 1.95$\mathrm{\pm}$1.32  & 2.09$\mathrm{\pm}$1.93  & 2.53$\mathrm{\pm}$2.01  & 1.39$\mathrm{\pm}$0.99  & 1.57$\mathrm{\pm}$0.99  & 1.55$\mathrm{\pm}$0.96  & 1.84$\mathrm{\pm}$1.76  \\ \hline 
5 & 7.48$\mathrm{\pm}$5.51 & 1.72$\mathrm{\pm}$1.6  & 2.0$\mathrm{\pm}$1.6 & 1.77$\mathrm{\pm}$1.53 & 2.07$\mathrm{\pm}$1.59  & 1.95$\mathrm{\pm}$2.10  & 2.06$\mathrm{\pm}$1.56  & 1.43$\mathrm{\pm}$1.31  & 1.62$\mathrm{\pm}$1.30  & 1.72$\mathrm{\pm}$1.28  & 1.80$\mathrm{\pm}$1.60  \\ \hline 
6 & 10.89$\mathrm{\pm}$6.9 & 1.49$\mathrm{\pm}$1.0  & 1.7$\mathrm{\pm}$1.0 & 1.29$\mathrm{\pm}$0.85 & 3.04$\mathrm{\pm}$2.73  & 5.16$\mathrm{\pm}$7.09  & 2.90$\mathrm{\pm}$1.70  & 2.26$\mathrm{\pm}$2.93  & 2.75$\mathrm{\pm}$2.91  & 2.02$\mathrm{\pm}$1.70  & 2.30$\mathrm{\pm}$3.78  \\ \hline 
7 & 11.03$\mathrm{\pm}$7.4 & 1.58$\mathrm{\pm}$1.2  & 1.9$\mathrm{\pm}$1.2 & 1.26$\mathrm{\pm}$1.09 & 3.41$\mathrm{\pm}$2.75  & 3.05$\mathrm{\pm}$3.01  & 3.60$\mathrm{\pm}$2.99  & 1.42$\mathrm{\pm}$1.16  & 2.34$\mathrm{\pm}$2.32  & 1.70$\mathrm{\pm}$1.03  & 1.91$\mathrm{\pm}$1.65  \\ \hline 
8 & 15.0$\mathrm{\pm}$9.01 & 2.11$\mathrm{\pm}$2.4  & 2.2$\mathrm{\pm}$2.3 & 1.87$\mathrm{\pm}$2.57 & 2.80$\mathrm{\pm}$2.46  & 6.48$\mathrm{\pm}$5.37  & 5.29$\mathrm{\pm}$5.52  & 3.13$\mathrm{\pm}$3.77  & 3.29$\mathrm{\pm}$4.32  & 2.64$\mathrm{\pm}$2.78  & 3.47$\mathrm{\pm}$5.00  \\ \hline 
9 & 7.92$\mathrm{\pm}$3.98 & 1.36$\mathrm{\pm}$0.7  & 1.6$\mathrm{\pm}$0.9 & 1.33$\mathrm{\pm}$0.98 & 2.18$\mathrm{\pm}$1.24  & 2.10$\mathrm{\pm}$1.66  & 2.38$\mathrm{\pm}$1.46  & 1.27$\mathrm{\pm}$0.94  & 1.86$\mathrm{\pm}$1.47  & 1.51$\mathrm{\pm}$0.94  & 1.47$\mathrm{\pm}$0.85  \\ \hline 
10 & 7.3$\mathrm{\pm}$6.35 & 1.43$\mathrm{\pm}$1.6  & 1.7$\mathrm{\pm}$1.2 & 1.14$\mathrm{\pm}$0.89 & 1.83$\mathrm{\pm}$1.36  & 2.09$\mathrm{\pm}$2.24  & 2.13$\mathrm{\pm}$1.88  & 1.93$\mathrm{\pm}$3.06  & 1.63$\mathrm{\pm}$1.29  & 1.79$\mathrm{\pm}$1.61  & 1.79$\mathrm{\pm}$2.24  \\ \hline 
\textbf{Mean} & \textbf{8.46$\boldsymbol{\mathrm{\pm}}$5.48} & \textbf{1.43$\boldsymbol{\mathrm{\pm}}$1.3 } & \textbf{1.66$\boldsymbol{\mathrm{\pm}}$1.14} & \textbf{1.32$\boldsymbol{\mathrm{\pm}}$1.24}\textbf{} & \textbf{2.17$\boldsymbol{\mathrm{\pm}}$1.89 } & \textbf{2.64$\boldsymbol{\mathrm{\pm}}$4.32 } & \textbf{2.50$\boldsymbol{\mathrm{\pm}}$1.16 } & \textbf{1.59$\boldsymbol{\mathrm{\pm}}$1.58 } & \textbf{1.86$\boldsymbol{\mathrm{\pm}}$2.12 } & \textbf{1.66$\boldsymbol{\mathrm{\pm}}$1.44 } & \textbf{1.83$\boldsymbol{\mathrm{\pm}}$2.35 } \\ \hline 
\end{tabular}
\end{table}
\bigbreak

\noindent 
\subsection{Brain}

Brain image registration has much wider options in databases than lung image registration. As a result, authors were not consistent on which database to use for training and testing and what metrics to use for validations. To facilitate benchmarking, we have listed a number of works on brain image registration in Table 9, which presents the datasets, the registration transformation model and the evaluation metrics. DSC of multiple ROI is the most commonly used evaluation metric. MI and surface distance measures are the next frequently used evaluation metrics.

\begin{table}
\footnotesize
\centering
\textbf{Table 9} Benchmark datasets and evaluation metrics used in brain registration

\begin{tabular*}{\textwidth}{c @{\extracolsep{\fill}} ccc} \\ \hline 
\textbf{References} & \textbf{Datasets} & \textbf{Transformation} & \textbf{Evaluation Metrics} \\ \hline 
\cite{RN15} & IXI & Deformable & TRE, MI \\ \hline 
\cite{RN18} & MindBoggle-101 & Deformable & DSC \\ \hline 
\cite{RN17} & BRATS, ALBERT & Affine & DSC, MI, SSIM, MSE \\ \hline 
\cite{RN11} & IBSR & Deformable & DSC, MI, NCC, SAD, DWT \\ \hline 
\cite{RN29} & LONI, LPBA40, IBSR, CUMC, MGH, IXI & Deformable & DSC, ASD \\ \hline 
\cite{RN36} & OASIS, ABIDE, ADHD, MCIC, PPMI, HABS, Harvard GSP & Deformable & DSC \\ \hline 
\cite{RN38} & ADNI & Deformable & DSC, SEN, PPV, ASD, HD \\ \hline 
\cite{RN46} & OASIS, IXI, ISLES & Rigid & MSE \\ \hline 
\cite{RN50} & IXI & Rigid & MAE of degree and translation \\ \hline 
\cite{RN59} & ADNI & Deformable & DSC \\ \hline 
\cite{RN74} & LONI, ADNI, IXI & Deformable & DSC, ASD \\ \hline 
\cite{RN79} & OASIS, IBIS, LPBA, IBSR, MGH, CUMC & Deformable & Target Overlap \\ \hline 
\cite{RN85} & LPBA & Deformable & TRE, JACC \\ \hline 
\cite{RN96} & IXI & Deformable & SSD, PSNR, SSIM \\ \hline 
\cite{RN103} & LONI, ADNI & Deformable & DSC \\ \hline 
\cite{RN106} & IXI, ALBERT & Deformable & DSC, JACC \\ \hline 
\end{tabular*}
\end{table}
\bigbreak

\noindent 
\section{Statistics}

\noindent After careful study of each category, it is important to step back and look at the whole picture. Out of the 150+ papers cited, more than half of the papers were aimed at direct transformation prediction using either supervised or unsupervised transformation prediction. The category of deep similarity-based methods accounts for 14\% of all methods while the category of GAN account for 10\% of all methods. Publications from the same group (Conference papers which were extended into journal papers) were counted only once if there were no substantial differences in content. One paper may belong to multiple categories. For example, unsupervised CNN method could use GAN generated loss for additional transformation regularization. Details percentages are shown in Fig.3. 
\begin{figure}
\centering
\noindent \includegraphics*[width=3.77in, height=1.81in, keepaspectratio=true]{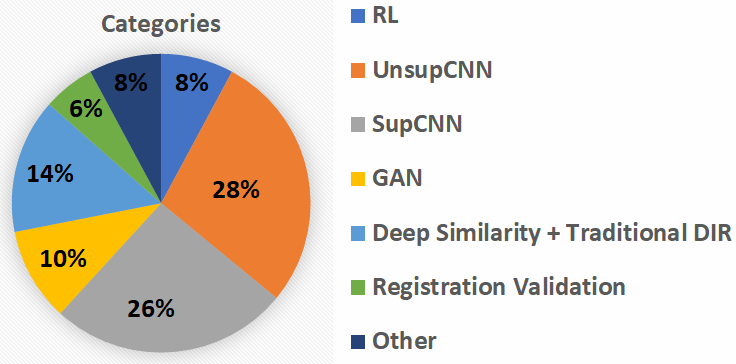}

Fig. 3. Percentage pie chart of different categories.

\end{figure}

Besides the number of papers, we have also analyzed the percentage distributions of many other attributes including input image pair dimension, transformation model, image domain, patch-based training, DL frameworks and ROI of the cited works. The percentage distributions were shown in Fig. 4. 60\% of the works were solving 3D-3D registration problems. The 2D-3D image registration works are mostly to register 3D-CT to 2D X-ray images for intraoperation image guidance. The percentages of the number of deformable, rigid and affine registration papers are 72\%, 19\% and 9\%, respectively. Most of the rigid registration papers are for intra-patient brain and spine alignment. There are more publications on unimodal than multi-modal image registration. Due to the superior performance of DL-based similarity measures to traditional similarity measures, the number of DL-based multi-modality image registration papers is increasing and accounts for 41\% of all the papers. Patch-based training was often adopted to save GPU memory. Fig.4 shows that 70\% of all works used whole image-based training. The 70\% includes not only 3D-3D but also 2D-3D and 2D-2D image registrations. Almost all 2D-2D registration used whole image-based training since 2D images are much less memory demanding than 3D images. Therefore, for 3D-3D image registration, there are roughly the same number of works that used whole image-based training and patch-based training. In terms of DL frameworks, Tensorflow is the leading framework which accounts for more than half of all papers. Pytorch is the second most popular DL framework which accounts for a quarter of all papers. Early works used Caffe and Theano, which was used less and less over the years as compared to Tensorflow and Pytorch. Theano has officially ceased development after version 1.0. The deep learning toolbox of Matlab is the least used framework perhaps due to licensing. In terms of the ROI, MR brain and CT lung are the most studied sites. Brain is the top registration target in all works. The reason for the wide adoption of brain include its clinical importance, its availability of public datasets and its relative simplicity of registration. 

\begin{figure}
\centering
\noindent \includegraphics*[width=6.50in, height=2.37in, keepaspectratio=true]{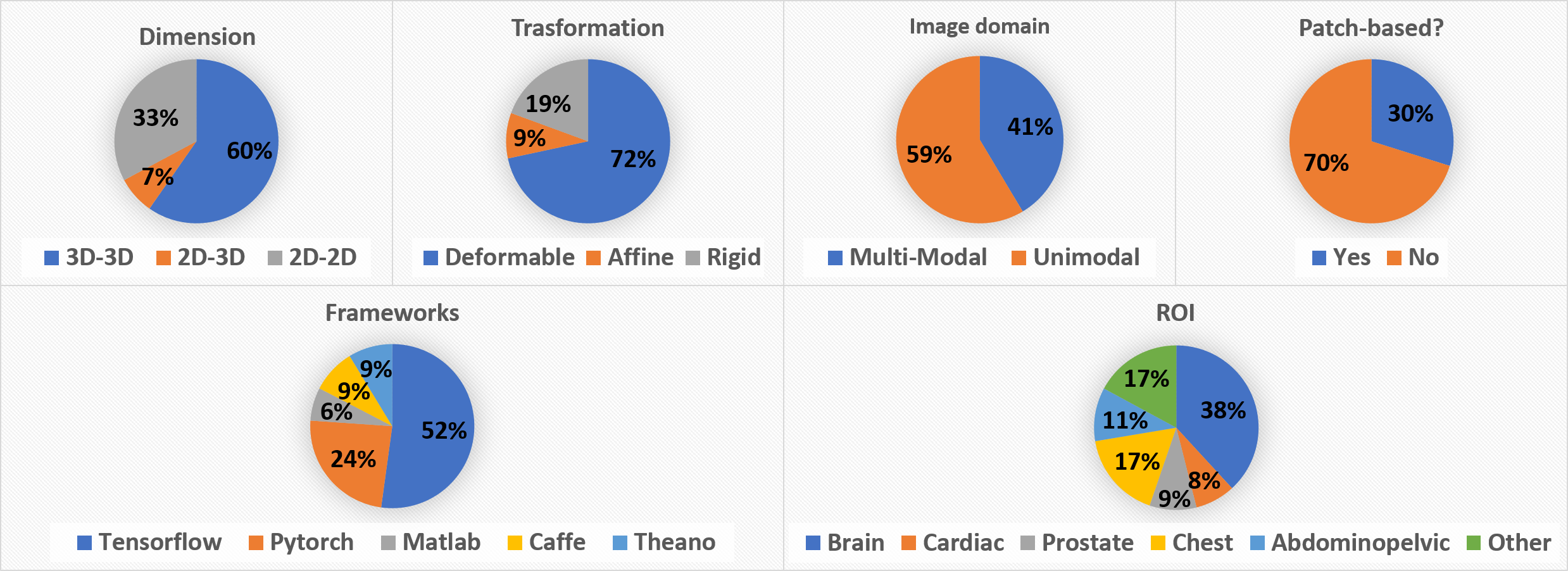}

Fig. 4 Percentage pie chart of various attributes of DL-based image registration methods.
\end{figure}

\noindent 

\noindent 
\section{Discussion}

Though image registration has been extensively studied, deep learning-based medical image registration is a relatively new research area. We have collected over 150 papers, most of which were published in the last 3 to 4 years. We generally classify these methods into seven non-exclusive categories. Many methods could be classified into multiple categories. For example, GAN was mostly used in combination with supervised or unsupervised transformation prediction methods as an auxiliary regularization or image pre-processing step. Supervised and unsupervised methods were combined for dual supervision in some works. Deep learning-based registration validation methods were included in this review because methods in this category often involve learning a deep similarity metric, therefore, could be used for image registration. RL and deep similarity-based methods are iterative whereas supervised and unsupervised based methods are non-iterative. For iterative methods, multiple works have reported that deep similarity metrics have superior performance to handcrafted intensity-based image similarity metric. For non-iterative methods, DL-based methods have yet to outperform traditional DIR methods. Take lung registration for example, the best performing DL-based methods are only comparable to the-state-of-art traditional DIR methods in terms of TRE. However, DL-based direct transformation methods are generally order of magnitude faster than traditional DIR methods. This is mainly due to the non-iterative nature and the powerful GPU utilized. A common feature that is used in both traditional DIR and DL-based methods is multi-scale strategy. Multi-scale registration could help the optimization avoid local maxima and allow large deformation registration. Regarding network generality, Fu \textit{et al.} and Jiang \textit{et al.} both showed that network trained using one set of datasets could be readily applied to an independent set of datasets given that the two image domains are close to each other. 

\noindent 
\subsection{Whole image-based vs. patch-based transformation prediction}

Whole image-based training and patch-based training have their own advantages and disadvantages. Due to limited GPU memory, the original images were often down-sampled to avoid memory overflow in whole image-based training. The down-sampling process could cause information loss and limit the registration accuracy. On the other hand, whole image training allows large inception field which enables registration of large deformations and mitigate the problem of local maxima in registration. Unless data augmentation is used, whole image-based training usually suffers from shortage of training datasets. On the contrary, patch-based training were not affected by the shortage of training datasets as much since many image patches could be sampled from the original images. In addition, patch-based training usually has better performance locally than whole-image based training. Recently, several groups combined whole-image training with patch-based training as a multi-scale approach for image registration \cite{RN192, RN21}. They have achieved promising results in terms of registration accuracy. One challenge with patch-based image registration is the patch fusion process, which stack many image patches to generate the final whole-image transformation. The patch fusion process could generate grid-like artifacts along the edges of the patches. One way to mitigate the problem is to use large patch overlap prior to patch fusion. However, it would make the inference process computationally inefficient. Another method is to use a non-parametric registration model for transformation prediction. One such example is LDDMM model used in QuickSilver \cite{RN79}. Instead of directly predicting final spatial transformation, QuickSilver predict the momentum of the LDDMM model. The LDDMM model can generate diffeomorphic spatial transformation without the need of smooth momentum predictions.

\noindent 
\subsection{Loss functions }

Despite large variations in details, loss function definitions of the cited works share many common features. Almost all loss function definitions consist of one or more combinations of the following six types of losses, which are 1) intensity-based image appearance loss, 2) deep similarity-based image appearance loss, 3) transformation smoothness constraint, 4) transformation physical fidelity loss, 5) transformation error loss with respect to ground truth transformation and 6) adversarial loss. Intensity-based image appearance loss includes SSD, MSE, MAE, MI, MIND, SSIM, CC and its variants. Deep similarity-based image appearance loss usually calculates the correlation between the learnt feature-based image descriptors. Transformation smoothness constraints usually involve the calculation of the first and second orders of spatial derivatives of predicted transformation. Transformation physical fidelity loss includes inverse consistency loss, negative Jacobian determinant loss, identity loss, anti-folding loss and so on. Transformation error loss was the error between predicted and ground truth transformations, which was only valid for supervised transformation prediction. Adversarial loss was the trainable network-based loss. Some auxiliary loss terms include the DSC loss of the anatomical labels or TRE loss of pre-selected landmark pairs.

\noindent 
\subsection{Challenges and Opportunities}

One of the most common challenges for supervised DL-based methods is the lack of training datasets with known transformations. This problem could be alleviated by various data augmentation methods. However, the data augmentation methods could introduce additional errors such as the bias of unrealistic artificial transformations and image domain shifts between training and testing stages. Several groups have demonstrated good generality of the trained network by applying them to datasets different from the training datasets. This inspired us to think that transfer learning maybe used to alleviate the problem of lack of training data. Surprisingly, transfer learning has not been used in medical image registration. For unsupervised methods, efforts were made to combine different kinds of regularization terms to constrain the predicted transformation. However, it is difficult to investigate the relative importance of each regularization term. Researchers are still trying to find an optimal set of transformation regularization terms that could help generate not only physically plausible but also physiologically realistic deformation field for a certain registration task. This is partially due to the lack of registration validation methods. Due to the unavailability of ground truth transformation between an image pair, it is hard to compare the performances of different registration methods. Therefore, registration validation methods are equally important as registration methods. We have observed an increased number of papers focusing on registration validation in 2019. More research on registration validation methods is desired in order to reliably evaluate the performances of different registration methods under different parametric configurations.

\noindent 
\subsection{Trends}

Judging from the statistics of the cited works, there is a clear trend of direct transformation prediction for fast image registration. So far, supervised and unsupervised transformation prediction methods are almost equally studied with close number of publications in either category. Either supervised or unsupervised methods have its own advantages and disadvantages. We speculate that more research will be focused on combining supervised and unsupervised methods in the future. GAN-based methods have gradually gaining popularity since GAN could be used to not only introduce additional regularizations but also perform image domain translation to cast multi-modal to unimodal image registration. We should see a steady growth of GAN-based medical image registration. New transformation regularization techniques have always been a hot topic due to the ill-posedness of the registration problem.

\noindent 
\bigbreak
{\bf Acknowledgements}

This research is supported in part by the National Cancer Institute of the National Institutes of Health under Award Number R01CA215718, and Dunwoody Golf Club Prostate Cancer Research Award, a philanthropic award provided by the Winship Cancer Institute of Emory University.

\noindent 
\bigbreak
{\bf Disclosures}

The authors declare no conflicts of interest.

\noindent 

\bibliographystyle{plainnat}  % needs package natbib
\bibliography{Manuscript_tex2}      

\end{document}